\documentclass[12pt,preprint]{aastex}  
\def\ergsec{\hbox{ergs s$^{-1}$}}

\slugcomment{accepted to the {\it Astrophysical Journal Letter} 2001 June 28}
\shorttitle{Globular cluster Population in NGC 1399}
\shortauthors{Angelini et al.}

\begin{document}

\title{The X-ray Globular Cluster Population in NGC 1399}
\author{Lorella Angelini\altaffilmark{1}, Michael Loewenstein\altaffilmark{2},
and Richard F. Mushotzky }
\affil{Laboratory for High Energy Astrophysics, NASA/GSFC, Code 662, 
Greenbelt, MD 20771}
\altaffiltext{1}{Also with the Universities Space Research Association}
\altaffiltext{2}{Also with the University of Maryland Department of Astronomy}
\email{angelini@davide.gsfc.nasa.gov}

\begin{abstract}

We report on the {\it Chandra} observations of the  elliptical galaxy NGC 1399,
concentrating on the X-ray sources identified with globular clusters (GCs).
A large fraction of the 2-10 keV X-ray emission 
in the $8'\times 8'$ {\it Chandra} image is resolved into point sources
with luminosities $\ge 5 \times 10^{37}$ \ergsec.
These sources are most likely Low Mass X-ray Binaries (LMXBs). 
In a region imaged by {\it HST} about 70\% of the X-ray sources 
are located within GCs. This association suggests that in giant elliptical 
galaxies luminous X-ray binaries preferentially form in GCs.
Many of the GC sources 
have super-Eddington luminosities (for an accreting neutron star) 
and their average luminosity is higher than the non-GC sources.
The X-ray spectral properties of both GC and non-GC sources are  
similar to those of LMXBs in our Galaxy. Two of the brightest 
sources, one of which is in a GC, have an ultra-soft spectrum, similar 
to that seen in the high state of black hole candidates.
The ``apparent'' super-Eddington luminosity in many cases may be 
due to multiple LMXB systems within individual GCs, but with some 
of the most extremely luminous systems containing massive black holes. 

\end{abstract}



\section{Introduction}

Early X-ray surveys in the 1970s revealed that the ratio of Low Mass
X-ray Binaries (LMXBs) to stellar mass is more than two orders of
magnitude higher for globular clusters (GCs) than it is for the rest of our
Galaxy (Clark 1975). This was surprising because globular clusters
contain only a small fraction, $\sim 1
\times 10^{-4}$, of the mass in our Galaxy.  
Yet, according to current estimates, they host 12
out of $\sim 130$ bright ($\ge 10^{36}$ \ergsec) Galactic X-ray
sources.  This overabundance of binaries led Fabian, Pringle, \& Rees
(1975) to propose that the LMXBs in GCs are formed via tidal capture of
neutron stars in close encounters with main-sequence or giant stars, a
mechanism that operates preferentially at the higher stellar densities
found in GCs.  As X-ray observatories have improved it has become
possible to study the X-ray properties of GCs associated with the
nearby galaxies. {\it ROSAT} observations of M31 revealed that the X-ray
properties of its GCs are similar to those observed in our Galaxy,
with a comparable fraction and luminosity distribution (Supper et
al. 1997).

Early X-ray observations of elliptical galaxies showed that the
dominant source of X-rays is  $\sim 1$~keV emission from hot
interstellar medium (ISM) (e.g., Forman, Jones, \& Tucker 1985).
{\it ASCA} spectral analysis of elliptical galaxies revealed the presence 
of a hard component, which was identified with an ensemble of LMXBs
(Matsumoto et al. 1997, Matsushita et al. 1994).  The LMXB origin was
confirmed by the $0.5''$ angular resolution {\it Chandra} images of 
two faint elliptical galaxies, NGC 1553
and NGC 4697, that resolved $\sim 60\% $ of the hard ($> 1$ keV) X-ray
emission into point sources 
(Sarazin, Irwin, \& Bregman 2000; Blanton, Sarazin, \& Irwin 2001).  
Further, in NGC 4697 at least $\sim 20 \% $
of the X-ray point sources were found to be associated with globular
clusters (Sarazin et al. 2001). The GC populations in elliptical galaxies, 
unlike our Galaxy and other late-type galaxies, show a bimodal distribution 
in optical colors and much higher number of GCs per galaxy luminosity 
(the specific frequency), with the highest number found in the central 
galaxies of clusters (Ashman \& Zepf 1992). 
NGC 1399 is a giant elliptical galaxy in the center of the Fornax Cluster.
It is particularly suitable for studying the X-ray properties of a GC 
population because it has four times the average elliptical galaxy 
GC specific frequency, and fifteen times that of a typical spiral 
galaxy. In this paper we present the X-ray properties of the GC population 
detected in the {\it Chandra} observations of NGC 1399.


\section{Observations and Data Reduction}

NGC 1399 was observed with the {\it Chandra} ACIS-S detector on 
January 18, 2000 for a total exposure of 55973 s.  The galaxy was positioned
at the center of the S3 chip, $\sim 2.3'$ from the aim point. 
We used the standard pipeline data processing, 
corrected the data for the gain and quantum efficiency maps, removed bad 
pixels, and applied a $1.5''$ aspect shift (see below).
Additional observations were obtained with the ACIS-I on October
18, 1999 ($\sim$ 3500 s) and January 19, 2000 (2000 s immediately following
the ACIS-S observation).  In this letter we concentrate on the S3 chip 
region that overlaps with observations taken with WFPC2 on {\it HST}
(Grillmair et al. 1999), which was used to identify X-ray sources. 
The properties of the entire sample of X-ray
sources detected in the {\it Chandra} images and the
characteristics of the diffuse emission are given in Mushotzky et
al. (2001, Paper II, in preparation).

The X-ray point sources were identified using a sliding cell
detection algorithm, as implemented in XIMAGE, on the cleaned event
file in the 0.3-10.0 keV energy band, with the detection threshold
set at $2 \sigma$. The central $20''$ is dominated in this energy band 
by the diffuse emission from the galaxy 
(see Fig 2 in Loewenstein et al. 2001) and is 
excluded from this analysis. A total of 214 sources were detected in 
the S3 chip of which 160 are above the $ 3 \sigma$ level. 
These sources comprise a large fraction of the 2-10 keV diffuse X-ray 
emission (Paper II) previously detected with {\it ASCA}.
The cosmic X-ray background is expected to contribute $\sim$ 20 objects 
across this region (Mushotzky et al. 2000).
Each source was visually inspected to ensure that it is a solid
detection. The 0.3--10 keV count rates were corrected for exposure,
vignetting, and point spread function. The source background was
evaluated in nearby regions.  The faintest source detected is $\sim
1.7 \times 10^{-4}$ cts~s$^{-1}$ -- corresponding to a luminosity of $5
\times 10^{37}$ \ergsec~ for a distance to NGC 1399 of 20.5 Mpc
(Merritt and Ferrarese 2001). 

\section{Source Identification with GCs and Their Characteristics}

To optically identify the {\it Chandra} X-ray sources we used the
{\it HST} WFPC2 images from Forbes et al. (1998).  The list of GCs (678
objects) derived from these images, along with their B and I band
magnitudes, were provided by C. Grillmair.  The {\it HST} image covers less
than 30\% of the $8' \times 8'$ S3 {\it Chandra} field of view and
includes 45 of the 214 X-ray sources detected in the S3 chip (Fig
1). 39 of these X-ray sources have a signal to noise ratio (SNR)
greater than $3 \sigma$.  One source is elongated in the X-ray image,
inconsistent with a single point source, and is excluded from further
analysis. 

The X-ray contours overlaid on the {\it HST} image reveal a systematic
offset in declination of about $1.5''$. A similar displacement was
seen comparing the October {\it Chandra} ACIS-I observation with the
January ACIS-S observation. Applying a declination shift of $1.5''$
to the {\it Chandra} data, the galaxy centroid measured in the {\it HST} 
image became consistent with that measured with {\it Chandra}, and the
ACIS-I and ACIS-S images come into alignment.  Following this
correction, we assumed that the absolute X-ray positions are accurate
to $\sim 0.5''$, the size of the {\it Chandra} point spread
function. For 26 X-ray sources the closest {\it HST} object -- all
identified as GCs -- is within $ 0.4 ''$ .
An additional 7 X-ray sources are $0.7''- 1''$ from
the nearest {\it HST} object, all GCs. The remaining 5 X-ray sources have positions more than $1 ''$ away from any {\it HST} object.  
We consider all 26 sources at 
distances $ < 0.4 ''$ from GCs as positive identifications. This
corresponds to $\sim 70\%$ of the (38) $3 \sigma$ sources. Among the 
$2 \sigma$ sources there are 3 additional possible GC identifications; 
however, we discuss the source characteristics only of the 
38 objects detected above $3 \sigma$.
If the $1.5''$ shift is not applied the number of matches 
within $ 0.5 ''$ is reduced to 1, demonstrating that chance coincidence 
is unlikely.
The GC X-ray sources are homogeneously distributed within the {\it HST} FOV
(Fig 1). The 38 objects have count rates that range
from $3\times 10^{-4}$ to $1\times 10^{-2}$ cts~s$^{-1}$.  
The GC X-ray sources are on average brighter than the unidentified 
sources. Three of the
sources, two of which are associated with GCs, are exceptionally
bright by factors $ > 5 $ above the average count rate.

We constructed a color-color diagram (Fig 2) by dividing the data into
3 energy bands: 0.3--1 keV, 1--2 keV and 2--10 keV. Only sources above
$5 \sigma$ ($> 42$ counts) are included in the plot.  There is a
distinct cluster of sources with a ``hard'' color, consistent with low
N$_h$ and power-law photon indices ranging from 1--2.  A few have softer
(power-law index 2--3) spectra. GC and unidentified sources occupy
identical regions in the color-color diagram.  Two of the three
brightest sources are very soft, with only one located in a
GC. The other very bright (GC) source is harder, located in the
densely populated region of the color-color diagram.
To determine the conversion factor to flux, composite spectra were 
obtained for the globular and non-globular samples, excluding the 
3 brightest sources that were individually fit. Absorbed power-law 
(with photon index $\sim 1.5$) and bremsstrahlung (with temperature 
$\sim 7.5$~keV) models fit both spectra with similar best-fit parameters 
and N$_{h}$ consistent with the Galactic value of
$1.3 \times 10^{20}$~cm$^{-2}$. Our adopted conversion factor is 
$6 \times 10^{-12}$ erg cm$^{-2}$ ct$^{-1}$, consistent within 
$ 10 \% $ for both models. Two of the brightest source spectra are well 
fit by power-laws of indices 3.4 and 2.7 (the former is in a GC) and 
absorption higher ($\sim 10^{21}$~cm$^{-2}$) than the Galactic value. 
The spectral parameters of the other brightest (GC) source are instead 
similar to those of the composite spectra.

Figure 3a compares the X-ray luminosity distribution for all 38 objects
with that of the X-ray GC sources. Most of the unidentified
sources are within the $ 1-2 \times 10^{38}$ \ergsec~ bin. 
Above $2\times 10^{38}$ \ergsec ~ the GC sources are clearly in excess
compared to the few unidentified sources. Excluding the 3 brightest sources, 
the average luminosity of the globular cluster X-ray sources is 
$3.2 \times 10^{38}$ \ergsec, compared to $ 2.1 \times 10^{38}$ for 
the unidentified sources. Including the brightest sources, the 
average increases by 2 and 1.5 for the globular and non-globular 
samples, respectively.

Only the 3 brightest sources have sufficient counts ($ > 500$) to
search for short term variability. Light curves with 320 and 3200
second bins of the 3 sources are consistent with a constant at
probability $ > 1 \times 10^{-2}$. Variability on a longer time scale, 
for the 3 brightest sources, was tested using the October 1999 observation. 
The predicted ACIS-I rate assuming the spectral parameters 
measured by ACIS-S are within $1 \sigma$ of the observed count 
rate in ACIS-I.

The average B and B-I of all X-ray globular sources above $3 \sigma$ 
are B$=23.2\pm 1.0 $ and B-I=$1.78\pm 0.28$ with 17 of 26 sources having
B-I larger than the average for the entire optical sample. 
The color (B-I) distribution for all clusters (see Fig 2 in Grillmair 
et al. 1999) is bimodal as seen in other elliptical galaxies. The GCs 
containing X-ray sources cover the full range of B-I, with a statistically 
marginal tendency to be redder than GCs in general. A weak correlation 
exists between the X-ray flux and B-I (redder GCs are fainter X-ray sources), 
that breaks down for the brightest sources. No correlation is seen 
between the X-ray flux and either B or I (Fig 3b), however there is a 
notable absence of X-ray detections in GCs fainter than 25$^{th}$ 
magnitude (Fig 3c). Overall, only a small percentage 
of optically luminous GCs contain luminous X-ray sources.  
The X-ray GC sources are on average $\sim 1.5$ magnitudes brighter 
in the optical than the average of all GCs in the {\it HST} FOV (Fig 3c). 
The KS test yields a low probability ($3.6 \times 10^{-6}$) 
that the X-ray globular magnitudes are drawn from the same distribution 
as the total globular sample. This increases to $4.3 \times 10^{-5}$ 
if the total sample is restricted to $1< B-I <2.5$.

\section{Discussion}

The {\it Chandra} observations of NGC 1399 presented here show that 
$\sim$70\% of the detected X-ray point sources, outside the central $20''$
of the galaxy, are associated with GCs. 
The average spectral properties of these sources are
similar to those of LMXBs in our Galaxy. 
The X-ray sources in GCs have on average a
higher luminosity compared to those not associated with GCs, with many
above $2\times 10^{38}$ \ergsec, the Eddington limit for spherical
accretion onto a 1.4 M$_{\odot}$ neutron star.  Estimates of the distance
to NGC 1399 range from $\sim 17-23$ Mpc (20.5 Mpc is used here;
Ferrarese et al. 2000). Even including a 30\% uncertainty in the
luminosity, the majority of sources remain above the
Eddington luminosity. These findings suggest that luminous LMXBs 
in giant elliptical galaxies are preferentially located in GCs.  This result
is quite remarkable considering that in our Galaxy (and in M31) only
10\% of LMXBs are in GCs -- with a typical luminosity of $10^{37}$
\ergsec, {\it well below the threshold luminosity, $5 \times 10^{37}$
\ergsec,  for detecting point sources in this {\it Chandra} exposure of
NGC 1399}. Moreover, there are no GC sources with $L_X > 10^{38}$ \ergsec~ 
in our Galaxy, nor in any of the $\sim 500$ GCs 
in M31 reported by Supper et al. (1997). Super-Eddington luminosities in GC 
sources are also seen in the faint elliptical galaxy NGC 4697 
(Sarazin et al. 2001).

The luminosity distribution (Fig 3a) of all sources in NGC 1399 seems to be
divided into 3 groups. The 3 extremely bright sources detected at 
luminosities greater than $ 10^{39}$ \ergsec~ form one group.  Of
these, two are identified with GCs and two have ultra-soft spectra.
A second, intermediate
luminosity, group of sources have luminosities between 2 and 10 $\times
10^{38}$ \ergsec.  Of these, 15 out of 18 are identified with GCs.  A
third group, with about half of the sources in GCs, have luminosities 
$ < 2 \times 10^{38}$ \ergsec~, with an average
of $1.4 \times 10^{38}$ \ergsec.

All the LMXBs found in GCs in our Galaxy have
luminosities between $ \sim 10^{36}$ \ergsec~ and $\sim 7 \times
10^{37}$ \ergsec~ with many of them transient sources (Hut et al. 1992).
The Galactic GCs tend to be, on average, less
luminous than the Galactic bulge or burst sources. Verbunt, van
Paradijs \& Elson (1984) showed that this difference cannot be
explained by the metallicity or mass of the secondary and concluded,
given the small number of sources, that the GC and Galactic sources are 
probably still drawn from the same luminosity distribution.  If the GC 
X-ray sources in 
NGC 1399 are formed by the tidal capture of a neutron star or the exchange 
collision mechanism  (Fabian, Pringle, \& Rees 1975; Hut, Murphy, \& Verbunt
1991), then these models must account for the large number of
super-Eddington luminosity sources.  One possibility is that some of these GCs
contain multiple LMXB systems. To explain the high-luminosity tail 
of the intermediate
luminosity group, requires either a modest number of LMXBs ($ < 5$) per GC at 
the Eddington luminosity or a larger number of lower luminosity systems 
resulting from a higher collision rate compared to Galactic clusters.

The higher luminosity sources above $10^{39}$ 
\ergsec, may form a distinct population.  The ultra-soft spectra observed 
in two of the brightest systems, one of which is in a GC, are similar
to those of high-state black holes seen in our Galaxy. 
The luminosities are then accounted for by fairly massive black 
holes ($ > 7 $ M$\odot$) accreting close to the Eddington limit. 
The last of the brightest 
GC sources has a power-law index of $\sim 1.5$, similar to the brighter 
sources in the intermediate luminosity group that also show harder 
color in the color-color plot. This spectral characteristic is typical 
of low-state black hole systems (like GS 2000+251, GS 1124-68) and,  
although this seems inconsistent with their high luminosity, similar 
spectral behavior at high luminosities has been seen from the black-hole 
transient GS2030+338 (Tanaka \& Lewin 1995) or from luminous 
X-ray sources in several nearby galaxies (Makishima et al. 2000).
It is particularly notable that
these sources may contain black holes, since no black hole candidates have 
previously been detected 
in GC sources in our Galaxy. It was suggested that 
this is because black holes formed in dense clusters are likely 
to be ejected from the cluster due to dynamical effects 
(Zwart \& McMillan 2000). In an intermediate density cluster, 
if a black hole survives, it may be able to form a LMXB 
(Kulkarni, Hut, \& McMillan 1993).  In our Galaxy such
systems are expected to be in quiescence most of the time -- with
occasional transient outbursts (White, Swank, \& Holt 1983).  If this
is the case, we expect to observe flux variations on a time-scale of a
month as the outburst decays. In our limited number of observations
this has not yet been observed, but given the lack of monitoring to
date, this is inconclusive.

The hard spectral component in elliptical galaxies seen with {\it ASCA} is
resolved into (mostly luminous) LMXBs. Both the high ratio of
hard-X-ray-to-optical luminosity and the high coincidence rate of
X-ray sources in globulars, can be attributed to the high globular
cluster specific frequency in NGC 1399. The GC bimodal B-I 
color distribution in NGC 1399 shows an unusually large fraction 
of red clusters. 
Whether the specific frequency or the abundance of red clusters
are the determining factors in the total number of LMXBs in ellipticals
and, indeed, in galaxies of all types (Sarazin et al. 2001), will be
determined as more {\it Chandra} galaxy observations are analyzed and 
compared.

The luminosity threshold of this {\it Chandra} observation of $5\times
10^{37}$ \ergsec~ exceeds most
of the luminosity range of the 12 GCs detected in our own Galaxy. 
The origin of the ``super-Eddington'' X-ray sources in the GCs in
NGC 1399 may simply be explained as multiple LMXBs within a single
GC, or they may be black holes. Clearly we are only sampling the tip of the
distribution of a much larger LMXB population. Understanding the relative
size of the GC and non-GC populations, how the GCs in NGC 1399 relates 
to that of our own Galaxy and models for their formation will require a 
factor of ten deeper exposure with {\it Chandra} to cover the entire 
luminosity range of LMXBs in GCs seen in our Galaxy.

Acknowledgements. We are very grateful to C. Grillmair for providing
the list of the globular positions and magnitudes, and to Eliot Malumuth
for assistance with {\it HST} images. We thank Frank Verbunt for useful
discussion and Nicholas White for many helpful comments.


\clearpage

\small
\begin{figure}
\figurenum{1}                            
\centerline{\includegraphics[scale=0.5,angle=-90]{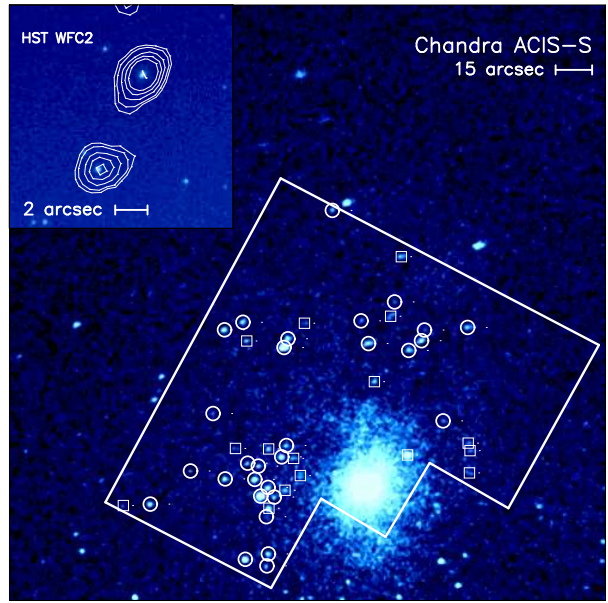}}
\caption{The 0.3-10 keV NGC 1399 ACIS-S image centered on the {\it HST}
pointing, smoothed with a Gaussian of about 0.8 arcsec.  The white
line marks the {\it HST}/WFPC2 FOV. The circles show the X-ray source
positions that are associated with globular clusters.  The squares are
the remaining sources. All 45 sources are marked; 38 are above 
$3\sigma$. The top left image is an example of the {\it Chandra}
2contours overlaid on the {\it HST} field.}
\end{figure}

\begin{figure} 
\figurenum{2}                            
\centerline{\includegraphics[scale=0.5,angle=-90]{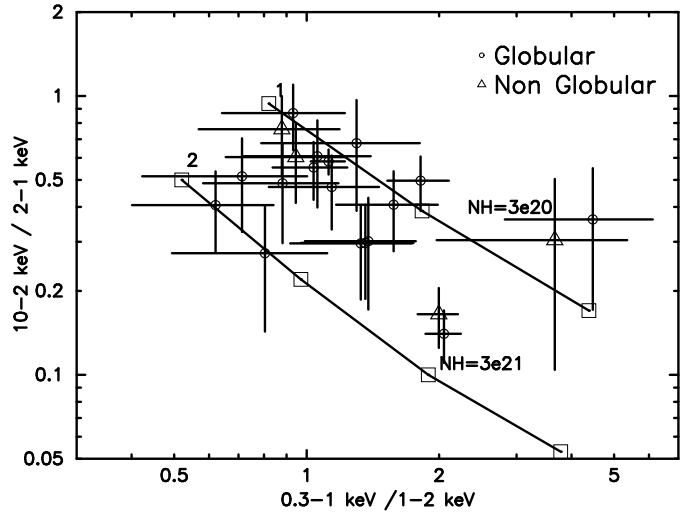}}
\caption{Color-color diagram of all sources detected above $ 5
\sigma$.  Most of the sources in the plot (80\%) are associated with
globular clusters (circles), since on average they are brighter than
those that are not (triangles).  The curves plotted represent
power-law models with constant N$_h$ with varying index.  At N$_h$~$ 3
\times 10^{20}$ the index varies from the top from 1-2-3.; at N$_h$~$
3 \times 10^{21} $ the index varies from 2-3-4-5.  Of the three
brightest sources two are located at $(X,Y)\sim(2,0.15)$, one (in a
GC) at $(1.12,0.58)$.  $5\sigma$ is chosen for clarity; globular and
non-globular clusters do not segregate if the threshold is reduced to
$ 3 \sigma$.}
\end{figure}

\begin{figure} 
\figurenum{3}                            
\centerline{\includegraphics[scale=0.53,angle=-90]{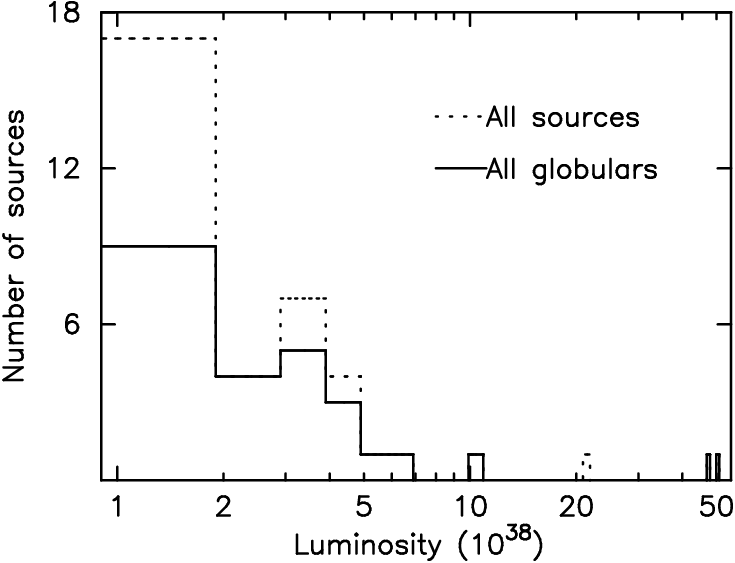}\includegraphics[scale=0.53,angle=-90]{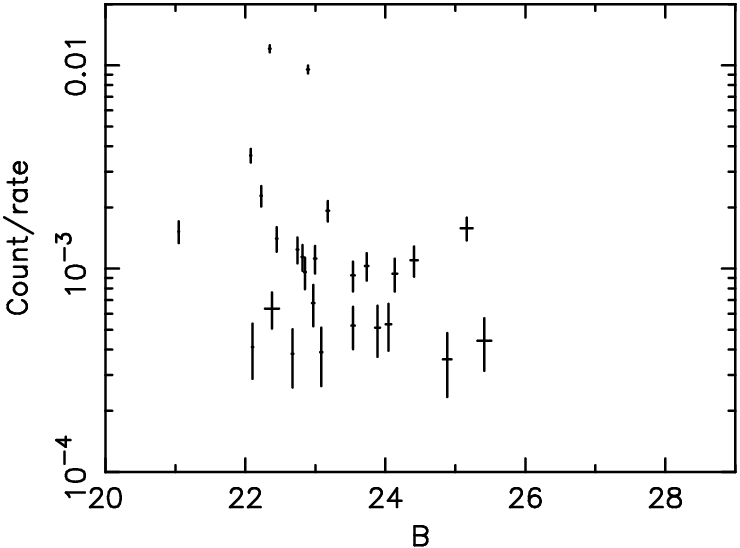}\includegraphics[scale=0.530,angle=-90]{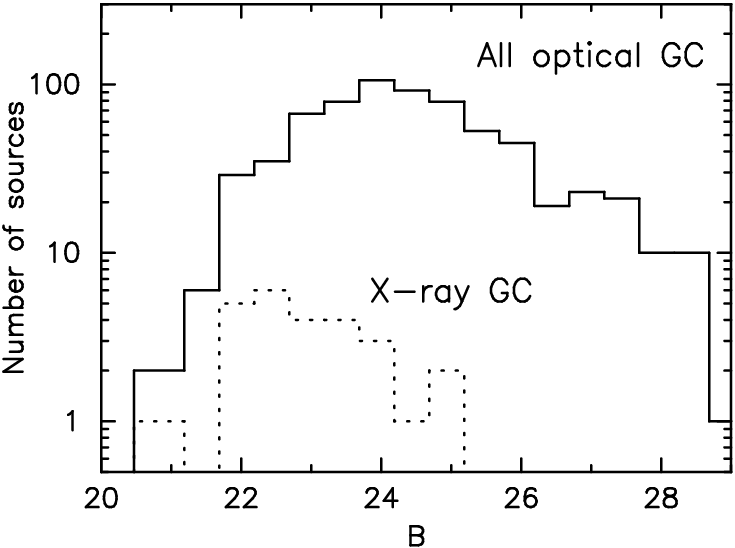}}
\caption{The luminosity distribution (left panel 3a) for all the X-ray
sources detected compared with the luminosity distribution of globular
X-ray sources.  The GC tend to be more luminous.  The X-ray count rate
does not show any obvious correlation with B (or I) magnitude as shown
in the middle panel (3b).  The B magnitude distribution (right panel 3c) of
the entire population of globulars detected in the {\it HST} field (solid
histogram) compared with that for globular sources with X-ray sources
detected at $>3 \sigma$.}
\end{figure}

\end{document}